\documentclass[twocolumn,prb,tightenlines,superscriptaddress,amsmath,amssymb,amsfonts,noeprint]{revtex4-2}
\usepackage{graphicx}
\usepackage{dcolumn}
\usepackage{bm}
\usepackage{float}
\usepackage{hyperref}
\usepackage[section]{placeins}
\usepackage{epstopdf}
\usepackage[table, dvipsnames]{xcolor}
\usepackage{newtxtext,newtxmath}

\DeclareMathOperator{\sech}{sech}

\begin{document}

\title{Localization-delocalization transition in non-integer-charged electron wave packets}
\author{Y. Yin}
\thanks{Author to  whom correspondence should be addressed}
\email{yin80@scu.edu.cn.}
\affiliation{Department of Physics,
  Sichuan University, Chengdu, Sichuan, 610065, China}
\date{\today}

\begin{abstract}
 
  We investigate the wave function of electrons and holes injected by voltage pulses with non-integer flux quantum. We
  find that the wave function can be delocalized in the time domain, which is measured by using the inverse
  participation ratio. As the flux approaches an integer multiples of flux quantum, the wave function can either remain
  delocalized or undergo a localization-delocalization transition. The former case occurs for the neutral electron-hole
  pairs, while the latter case occurs for electrons and holes which are injected individually. We perform the
  finite-size scaling analysis of the inverse participation ratio to further clarify the nature of the
  localization-delocalization transition. The scaling function and correlation length are determined numerically by the
  data collapse method. We find that the localization-delocalization transition is universal in the sense that the
  scaling function and correlation length is insensitive to the pulse profile when the pulse is sufficiently sharp. In
  contrast, they exhibit quantitatively different behavior for the Lorentzian pulse, indicating the corresponding
  localization-delocalization transition belongs to a different universality class.
  
\end{abstract}


\maketitle

%
%

\section{Introduction}
\label{sec1}

In a ballistic conductor, an electron can be injected coherently from a reservoir by using a time-dependent driven
voltage \cite{keeling_2006_minim, dubois_2013_minim, gabelli_2013_shapin, jullien_2014_quant, glattli_2016_levit,
  baeuerle_2018_coher, bisognin_2019_quant}. This provides a simply but feasible way to achieve the on-demand charge
injection. Although an electron always carries a negative elementary charge, the injected current can carry arbitrary
non-integer charges due to different mechanisms. A well-known mechanism is the charge fractionalization in the integer
quantum hall edges \cite{safi_1995_trans, pham_2000_fract, imura_2002_conduc, steinberg_2007_charg, hur_2008_charg}. Due
to electron-electron interaction, an electron wave packet injected in one edge can be split into several fragments, each
carrying a fractional of the electron's charge \cite{bocquillon_2013_separ, wahl_2014_inter}. The wave packet fragments
remain localized and well-separated in space and time \cite{kamata_2014_fract, perfetto_2014_time,
  hashisaka_2017_wavef}. This makes them behave like individual elementary excitations with non-integer charges
\cite{berg_2009_fract, inoue_2014_charg}. Alternatively, the non-integer charge can also be injected due to the
scattering. If an electron wave packet is transmitted into the conductor with the probability $T < 1$, a total amount of
$Te$ charges are injected into the conductor. In this case, the non-integer charges are attributed to a mixing between
one electron state and a vacuum state. They are simply quantum averages, which are always accompanied by non-vanishing
fluctuations.

The non-integer charges can also be injected in the absence of both interaction and scattering. This can be done by
applying a voltage pulse on the reservoir, which carries non-integer multiples of flux quantum \cite{dubois_2013_integ,
  hofer_2014_mach}. In this case, there is no signature of well-defined non-integer charges, indicating that they are
also quantum averages. Despite the simple situation, the mechanism of the non-integer charge injection is not fully
understood. The understanding is mainly hindered by the dynamical orthogonality catastrophe \cite{levitov_1996_elect,
  glattli_2017_pseud, yue_2019_normal}. Due to this phenomenon, a single pulse with non-integer flux quantum alway
injects a divergent number of neutral electron-hole (\textit{eh}) pairs into the conductor \cite{levitov_1996_elect,
  sherkunov_2010_quant, knap_2012_time}. This prevents one from direct characterizing the quantum states of electrons
and holes. This problem can be avoided in a specific case when the charge injection is driven by a Lorentzian pulse with
one-half flux quantum \cite{moskalets_2016_fract}. In this case, one can construct a single-particle excitation, which
is orthogonal to the multi-particle state of neutral \textit{eh} pairs. It is essentially a mixed state, which is
composed of an electron wave packet and a vacuum state, each with associated probability $1/2$. From this result, it
seems that the non-integer charge injection is due to the classical mixing, which is similar in spirit to the case of
scattering. However, it is not clear how the classical mixing can arise in the absence of interaction and scattering. In
fact, when the injection is driven by periodic pulses, the non-integer charge injection can be understood in an
alternatively way. In a previous work, the author has found that the time width of single-electron wave packets can be
larger than the period of the driving pulses when each pulse carries a non-integer flux quantum
\cite{yue_2021_quasip}. In this case, the charges injected per period are measured over a limit time region, which is
smaller than the size of the electron wave function. This leads to the non-integer charge injection per period, which
are always accompanied with non-vanishing quantum fluctuations.

The purpose of this paper is to show that this picture also holds when the charge injection is driven by a single
voltage pulse with a non-integer flux quantum. This indicates that the corresponding wave function are essentially
delocalized in the time domain. To show this, we consider two successive pulses with the same shape but opposite
signs. Given a fixed pulse width, the single pulse limit can be approached by increasing the time interval between the
two pulses to infinity. This provides a controllable way to circumvent the dynamical orthogonality.

We first identify the wave functions of electrons and holes which are injected individually by a single voltage
pulse. The extent of the wave function in the time domain is then analyzed quantitatively by using the inverse
participation ratio (IPR). As the flux of the pulse drops from one flux quantum to zero, we find that the wave function
can undergo a localization-delocalization (LD) transition, which can be seen from the length-scale dependence of the
IPR. In contrast, the wave functions of neutral \textit{eh} pairs remain localized or delocalized as the flux varies.

We then perform the finite-size scaling analysis to investigate the LD transition in details. We find that the
corresponding IPR can be described by a single parameter scaling function. The scaling function and the corresponding
critical exponent can be extracted numerically by using the data collapse method. We find that the LD transition is
universal in the sense that the scaling function and the critical exponent and is insensitive to the pulse profile when
the pulse is sufficiently sharp. In contrast, they can show a quantitatively different behavior for the Lorentzian
pulse, indicating the LD transition for the Lorentzian pulse belongs to a different universality class.

The paper is organized as follows. In Sec.~\ref{sec2}, we present our model for the charge injection and show how to
extract the wave function from the scattering matrix. In Sec.~\ref{sec3}, we introduce the IPR to distinguish the
localized and delocalized wave functions. In Sec.~\ref{sec4}, we perform the finite-size scaling analysis of the IPR to
investigate the LD transition. We summarize in Sec.~\ref{sec5}.

\section{Model and formalism}
\label{sec2}

\begin{figure}
  \centering
  \includegraphics[width=8.5cm]{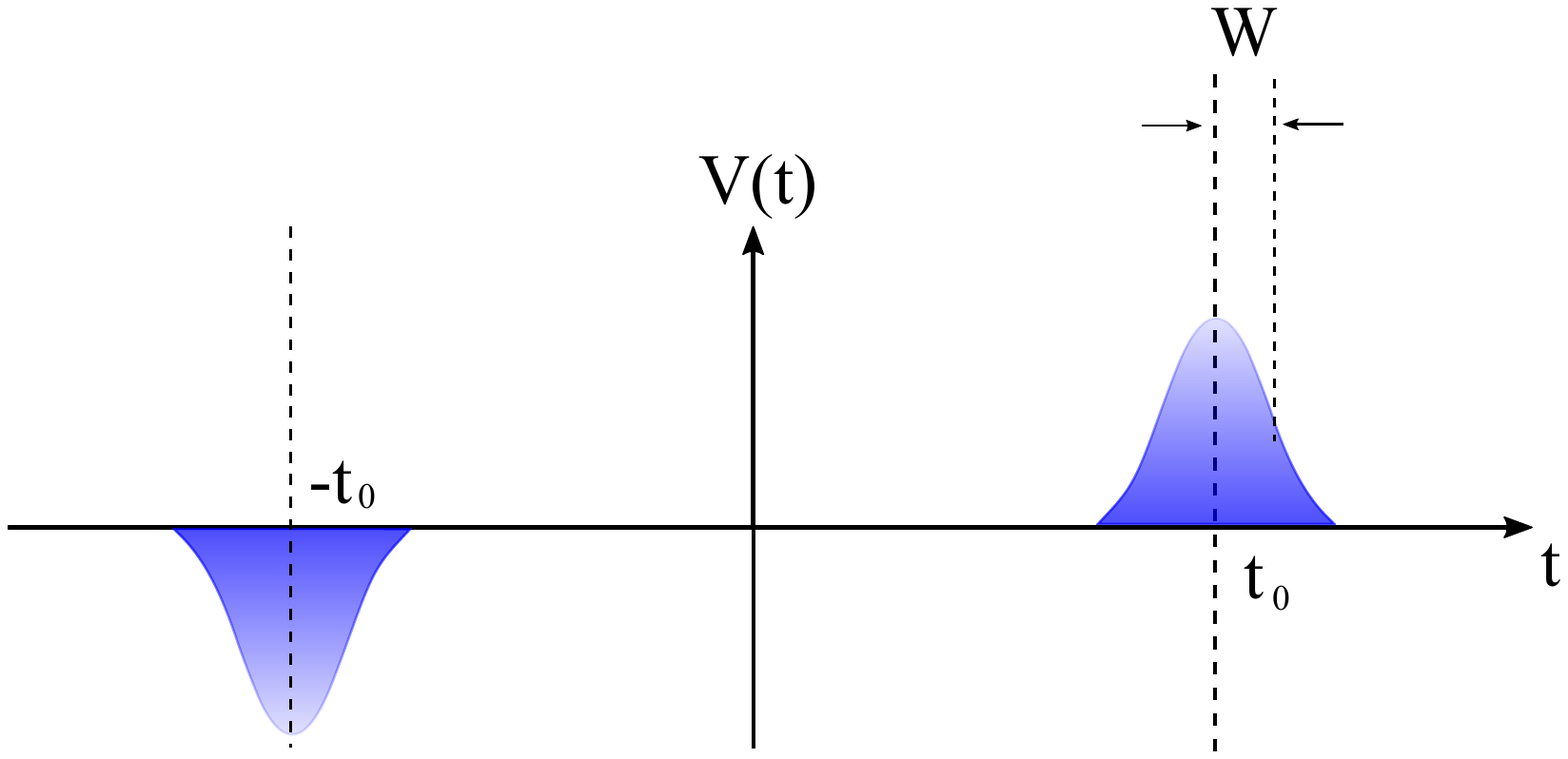}
  \caption{ Schematic of the voltage pulses. The two voltage pulses have the same shape but opposite signs. The width of
    each pulse is $W$, which are separated by the time interval $2t_0$.}
  \label{fig1}
\end{figure}

We consider the charge injection from a reservoir into a single-mode quantum conductor. We assume the injection is
driven by a time-dependent voltage $V(t)$ applied on the electrode. We choose the driving voltage $V(t)$ of the form
\begin{eqnarray}
  V(t) = V_p(t_0+t) - V_p(t_0-t),
  \label{s2:eq10}
\end{eqnarray}
indicating that it is composed of two successive pulses with the same shape but opposite signs, as illustrated in
Fig.~\ref{fig1}. The width of each pulse can be characterized by the half width at half maximum (HWHM) $W$, which are
separated by a time interval $2t_0$, as illustrated in Fig.~\ref{fig1}. The strength of the pulse can be described by
the flux $\varphi = (e/h) \int^{+\infty}_{-\infty} V_p(t) dt$, which is the Faraday flux of the voltage pulse normalized
to the flux quantum $h/2e$ with $h$ being the Planck constant and $e$ is the electron charge. Throughout this paper, we
set $e=h=W=1$. The current induced by the emitted electrons can be related to the driving voltage as
$I(t)= V_p(t_0+t) - V_p(t_0-t)$. The charge $Q$ injected by each pulse is equal to
\begin{equation}
  Q = \pm \int^{+\infty}_{-\infty} dt V_p(t) = \pm \varphi.
\end{equation}

The electron emission in this system can be fully characterized by the scattering matrix
\begin{eqnarray}
  b(E) = \int^{+\infty}_{-\infty} dE' S(E - E') a(E'),
  \label{s2:eq20}
\end{eqnarray}
where $a(E)$ and $b(E)$ represent the electron annihilation operators in the electrode and the quantum conductor,
respectively. The matrix element $S(E - E')$ is only the function of the energy difference $E-E'$, which can be related
to the voltage pulse $V(t)$ as
\begin{eqnarray}
  S(E - E') = \int^{+\infty}_{-\infty} dt e^{2\pi i (E-E') t - i\phi(t)},
  \label{s2:eq30}
\end{eqnarray}
with $\phi(t) = \int^t_{-\infty} V(\tau) d\tau$ being the scattering phase due to the driving pulses. Note that for the
voltage pulse we considered here, the scattering matrix is symmetric, {\em i.e.}, $S(E - E') = S(E' - E)$, since we have
$V(t) = -V(-t)$ from Eq.~\eqref{s2:eq10}.

In our previous work, we have shown that the many-body state of the injected electrons can be obtained from the polar
decomposition of the scattering matrix \cite{yin_2019_quasip}. For the symmetric scattering matrix we considered here,
the decomposition can be obtained by solving the following eigen equations:
\begin{widetext}
  \begin{eqnarray}
    \int^{+\infty}_{0} dE' S(E-E') [\psi^e_k(E')]^{\ast} = \sqrt{1-p_k} \psi^e_k(E) + i \sqrt{p_k} \psi^h_k(E), \label{s2:eq40-1}\\
    \int^{0}_{-\infty} dE' S(E-E') [\psi^h_k(E')]^{\ast} = \sqrt{1-p_k} \psi^h_k(E) + i \sqrt{p_k} \psi^e_k(E). \label{s2:eq40-2}
  \end{eqnarray}
\end{widetext}
The corresponding many-body state can be expressed as
\begin{equation}
  | \Psi \rangle = \prod_{k=1, 2, 3, \dots} \Big[ \sqrt{1 - p_k} + i \sqrt{p_k} B^{\dagger}_e(k) B^{\dagger}_h(k) \Big] | F \rangle, \label{s2:eq50}
\end{equation}
where $|F\rangle$ represents the Fermi sea and $B^{\dagger}_e(k)$[$B^{\dagger}_h(k)$] represents the creation operator
for the electron[hole] component of the \textit{eh} pairs. They can be related to the polar decomposition Eqs.~(\ref{s2:eq40-1})
and (\ref{s2:eq40-2}) as
\begin{eqnarray}
  B^{\dagger}_e(k) & = & \int^{+\infty}_0 dE \psi^e_k(E) a^{\dagger}(E), \\
  B^{\dagger}_h(k) & = & \int^0_{-\infty} dE \psi^h_k(E) a(E).
                         \label{s2:eq60}
\end{eqnarray}

Due to the symmetric of the scattering matrix $S(E - E') = S(E' - E)$, Eqs.~(\ref{s2:eq40-1}) and~(\ref{s2:eq40-2}) can
be further combined into a single equation for $E>0$:
\begin{equation}
  \int^{+\infty}_{0} dE' S(E+E') \psi^{\ast}_{k}(E') = i \sigma \sqrt{p_k} \psi_k(E),
  \label{s2:eq70}
\end{equation}
with $\sigma = \pm 1$. The wave function of the electron and hole can be obtained as
$\psi^e_k(E) = \sigma \psi^h_k(-E) = \psi_k(E)$. In the numerical calculation, Eq.~(\ref{s2:eq70}) can be solved in the
energy domain by using the singular value decomposition.

Intuitively, one expect that the individual electrons (holes) and \textit{eh} pairs injected by the two pulses are
well-separated as $t_0 \gg 1$ and all the wave functions are well-localized around either $t=-t_0$ or $t=t_0$: The
former are solely due to the negative pulse, while the latter are solely due to the positive pulse. However, in the next
section, we shall show that this is not true when the pulse carries non-integer charges.

\section{Localized and Delocalized states}
\label{sec3}

\begin{figure}
  \centering
  \includegraphics{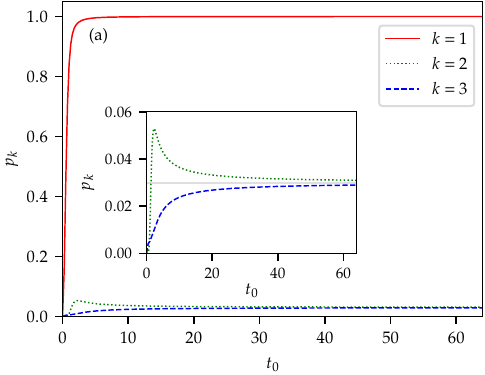}
  \includegraphics{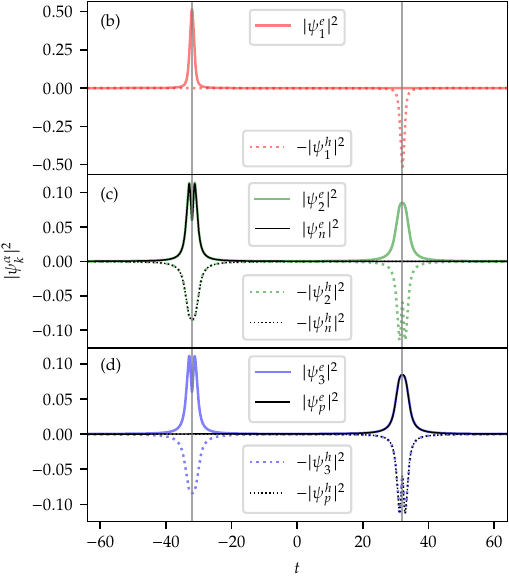}
  \caption{ (a) The excitation probabilities $p_k$ as a function of $t_0$ for $\varphi = 1.0$. The inset shows the
    zoom-in for $p_k < 0.06$. The probabilities $p_k$ of other \textit{eh} pairs are too small to be seen from the
    figure. (b-d) Modulus square of wave functions for electrons ($|\psi^e_k|^2$) and holes ($-|\psi^h_k|^2$),
    corresponding to $\varphi = 1.0$ and $t_0=32$. }
  \label{fig2}
\end{figure}

To demonstrate this, let us first consider a pair of Gaussian pulses with $\varphi=1.0$. The typical behavior of the
excitation probabilities $p_k$ as a function of $t_0$ are plotted in Fig.~\ref{fig2}(a). In this case, the
excitation is dominated by the first three \textit{eh} pairs. Their excitation probabilities are sensitive to $t_0$ when
$t_0$ is small, but become $t_0$-independent in the limit $t_0 \to +\infty$. One can see that $p_1$ approaches $1.0$,
while both $p_2$ and $p_3$ approach $0.03$.

The typical behavior of their wave functions in the large $t_0$ limit are demonstrated by their modulus square in
Fig.~\ref{fig2}(b--d), corresponding to $t_0=32$. Note that one always has
$\left|\psi^e_k(t)\right|^2 = \left|\psi^h_k(-t)\right|^2$, since the wave functions for electron and holes satisfy
$\psi^e_k(E) = \sigma \psi^h_k(-E)$ from Eq.~(\ref{s2:eq70}). For the first \textit{eh} pair ($k=1$), the electron and
hole wave functions are well-separated in the time domain, which are localized either around $t=-t_0$ or $t=t_0$. This
is shown in Fig.~\ref{fig2}(b), where the red solid and dotted curves correspond to the wave function of the
electron [$|\psi^e_1|^2$] and hole [$-|\psi^h_1|^2$], respectively. This indicates that the first \textit{eh} pair do
not belong to the neutral cloud of \textit{eh} pairs excited by a single pulse. It is composed of the single electron
and hole, which are injected by the two pulses individually.

The wave functions of the second [$\pm |\psi^{e/h}_2|^2$] and third [$\pm |\psi^{e/h}_3|^2$] \textit{eh} pairs are
demonstrated in Fig.~\ref{fig2}(c) and (d) by the green and blue curves, respectively. The electron and hole wave
functions are strongly overlapped in the time domain, indicating that they correspond to neutral \textit{eh}
pairs. Indeed, they are bonding-like and anti-bonding-like states built from two localized \textit{eh} pairs. In the
limit $t_0 \to +\infty$, we find that the localized wave functions can be constructed as
\begin{eqnarray}
  \psi^{e/h}_n(t) & = & \frac{ \psi^{e/h}_2(t) + i \psi^{e/h}_3(t) }{\sqrt{2}}, \label{s3:eq10-1}\\
  \psi^{e/h}_p(t) & = & \frac{ \psi^{e/h}_2(t) - i \psi^{e/h}_3(t) }{\sqrt{2}}. \label{s3:eq10-2}
\end{eqnarray}
They are plotted by the black solid and dashed curves in Fig.~\ref{fig2}(c) and (d). In this case, $\psi^e_n$ and
$\psi^h_n$ in Fig.~\ref{fig2}(c) forms a neutral \textit{eh} pair around $t=-t_0$, while $\psi^e_p$ and $\psi^h_p$
in Fig.~\ref{fig2}(d) forms a neutral \textit{eh} pair around $t=t_0$. In doing so, we have separated the wave
function of individual electron and hole ($k=1$) from the ones of the neutral \textit{eh} pairs ($k=2$ and $3$). Both of
them are localized wave functions for $\varphi=1$.

\begin{figure}
  \centering
  \includegraphics{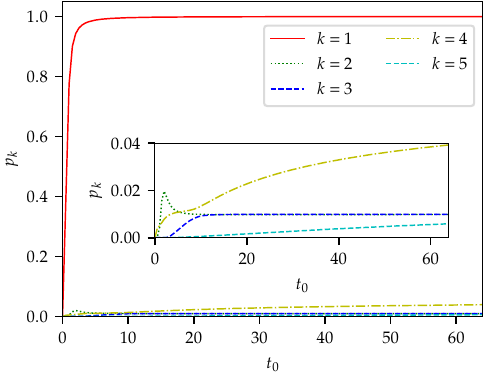}
  \caption{ The excitation probabilities $p_k$ as a function of $t_0$ for $\varphi = 0.9$. The inset shows the zoom-in
    for $p_k < 0.04$. The probabilities $p_k$ of other \textit{eh} pairs are too small to be seen from the figure.}
  \label{fig3}
\end{figure}

However, the scenario is quite different for pulses with non-integer flux quantum. To show this, let us consider a pair
of Gaussian pulses with $\varphi=0.9$. The typical behavior of the excitation probabilities demonstrated in
Fig.~\ref{fig3}. One can identify five excitation probabilities in this case, which have two different
$t_0$-dependence in the limit $t_0 \to +\infty$. The first three ones ($p_1$, $p_2$ and $p_3$) are insensitive to $t_0$
in this limit: While $p_1$ approaches $1.0$, $p_2$ and $p_3$ approach the same value $0.01$. In contrast, the last two
ones ($p_4$ and $p_5$) always increase with $t_0$.

\begin{figure}
  \centering
  \includegraphics{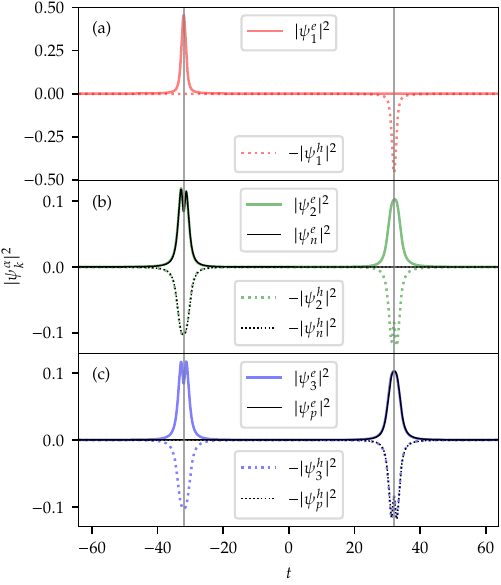}
  \caption{ Modulus square of localized wave functions for the electron ($|\psi^e_k|^2$) and hole ($-|\psi^h_k|^2$),
    corresponding to $\varphi = 0.9$ and $t_0=32$. }
  \label{fig4}
\end{figure}

The two different $t_0$-dependence of $p_k$ indicate two different nature of the \textit{eh} pairs, which can be seen
intuitively from their wave functions. The typical behavior of the wave functions for the first three ones ($p_1$, $p_2$
and $p_3$) are shown in Fig.~\ref{fig4}, corresponding to $t_0=32$. One can see that the wave functions of the
first \textit{eh} pair ($k=1$) are concentrated around either $t = - t_0$ or $t = t_0$, indicating that they are still
composed of localized states. One can also construct the localized wave functions $\psi^{e/h}_n$ and $\psi^{e/h}_p$ from
$\psi^{e/h}_2$ and $\psi^{e/h}_3$ by using Eqs.~(\ref{s3:eq10-1}) and~(\ref{s3:eq10-2}), indicating that they are still
bonding and anti-bonding states. In fact, by comparing Fig.~\ref{fig4}(a-c) to Fig.~\ref{fig2}(b-d), one finds
that these wave functions exhibit quite similar profiles, which are all localized in the time domain.

\begin{figure}
  \centering
  \includegraphics{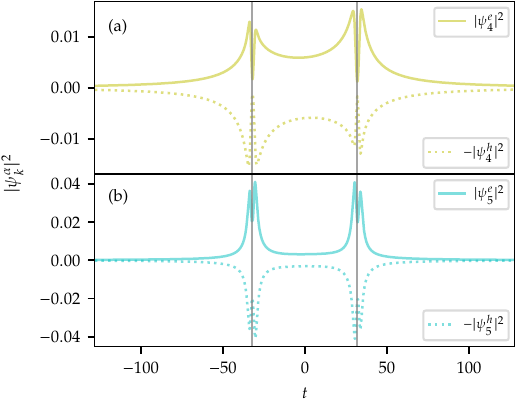}
  \caption{ Modulus square of delocalized wave functions for the electron ($|\psi^e_k|^2$) and hole ($-|\psi^h_k|^2$),
    corresponding to $\varphi = 0.9$ and $t_0=32$. }
  \label{fig5}
\end{figure}

In contrast, the last two ones $p_4$ and $p_5$ correspond to a new type of \textit{eh} pairs, which has no counterparts
for $\varphi = 1.0$. The typical behavior of their wave functions are demonstrated in Fig.~\ref{fig5}(a) and (b),
corresponding to $t_0=32$. One can see that the wave functions of electrons and holes are largely overlapped in the time
domain, indicating that they are also neutral \textit{eh} pairs. However, their wave functions exhibit much broad
profiles, which distribute more or less evenly between $-t_0$ and $t_0$ and drops slowly to zero outside this
region. This suggests that they are essentially delocalized states. The two voltage pulses act effectively as two
barriers in the time domain, which confines the delocalized wave functions inside a box with width proportional to
$t_0$.

\begin{figure}
  \centering
  \includegraphics{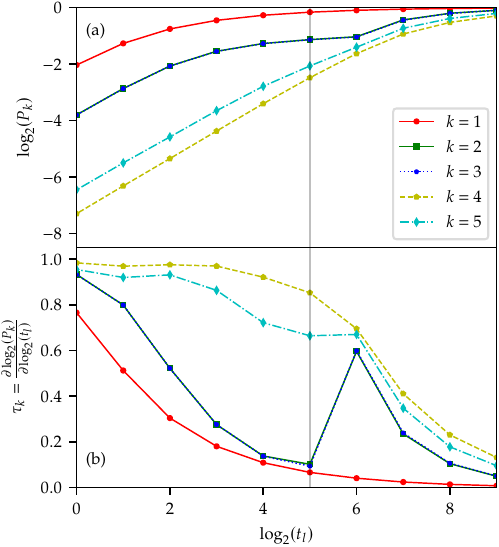}
  \caption{ (a) The logarithm of the IPR $\log_2(P_k)$ as a function of logarithm of the box size $\log_2(t_l)$ for
    $t_0=32$ and $\varphi=0.9$. The black vertical solid line corresponds to $t_l=t_0$. (b) The exponent $\tau_k$
    estimated from the slope of the logarithm plot. }
  \label{fig6}
\end{figure}

The delocalization nature can be characterized in a more concrete way from the length-scale dependence of the inverse
participation ratios (IPR), which has been extensively used in characterizing the localized/delocalized states in the
Anderson transition \cite{wegner_1980_inver, aoki_1983_critic, castellani_1986_multif, janssen_1994_multif,
  janssen_1998_statis, yakubo_1998_finit, brndiar_2006_univer, evers_2008_ander_trans, murphy_2011_gener,
  iyer_2013_many, tikhonov_2016_ander}. The length-scale dependence can be investigated by varying the system size, or
by dividing the system into integer number of small boxes of linear size $t_l$ and varying $t_l$ while keeping the
system size fixed \cite{yakubo_1998_finit, rodriguez_2008_multif}. As the electron and hole wave functions have the same
profile, we shall concentrate on the electron wave function and suppress the superscript ($e$) for clarify. For the
electron wave function $\psi_k(t)$ in the time domain, the IPR for a fixed $t_0$ can be defined as
\begin{equation}
  P_k = \frac{1}{t_l} \int_{\rm{box\: origins}}  \sum_{\rm{box}(t_l)} \Big( \int_{t \in \rm{box}(t_l)} \left| \psi_k \right|^2  \Big)^2, \label{s3:eq20}
\end{equation}
where we assume the wave function is normalized in the whole time domain
$\int^{+t_{\rm max}/2}_{-t_{\rm max}/2} dt \left| \psi_k \right|^2 = 1$ with $t_{\rm max} \to +\infty$. The symbol
$\sum_{\rm{box}(t_l)}$ represents the summation over small boxes with a linear size $t_l$ into which one divides the
whole time domain $[-t_{\rm max}/2, +t_{\rm max}/2]$. The choice of the box origins is arbitrary, which gives different
values of IPR. Following Refs.~\cite{rodriguez_2008_multif, rodriguez_2009_optim}, the IPR is further averaged over
different box origins ($\frac{1}{t_l} \int_{\rm{box\: origins}} \dots$) to avoid this ambiguity.

When $t_l$ is far from any characteristic time scales, the IPR $P_k$ scales with the box size $t_l$ as a power-law
\begin{equation}
  P_k \sim t^{\tau_k}_l.
  \label{s3:eq30}
\end{equation}
For the localized state, one has $\tau_k=0$ when $t_l$ is much larger than the localization length, while $\tau_k>0$
when $t_l$ is much smaller than the localization length. For the delocalized state which corresponds to an infinite
localization length, one always has $\tau_k>0$. In the time domain, which is a one-dimensional system, the upper bound
of $\tau_k$ is $1$, corresponding to the wave function which is uniformly extended in the time domain. For $0<\tau_k<1$,
the wave function is partially extended, which exhibit a self-similar fractal structure. This provides a quantitative
way to distinguish the localized and delocalized states.

The typical behavior of the IPR are demonstrated in Fig.~\ref{fig6}(a), corresponding to $t_0=32$ and $\varphi =
0.9$. In the calculation, we choose $t_{\rm max}=8192$, which is large enough to provide a compact support of all the
wave functions. The slope $\log_2(P_k)/\log_2(t_l)$ obtained from the numerical derivative is plotted in
Fig.~\ref{fig6}(b), from which one can estimate the exponent $\tau_k$ from Eq.~(\ref{s3:eq30}). One first notice that
all the IPRs approach $1.0$ when $t_l$ is much larger then $t_0$ (the black vertical line). This suggests that the two
voltage pulses act as soft barriers, which confines all the wave functions into a box comparable to $t_0$. 

In contrast, the IPRs show two distinctly different dependences on $t_l$ for $t_l < t_0$: As $t_l$ increases to $t_0$,
the IPR $P_1$, $P_2$ and $P_3$ approach constants and hence the corresponding exponent $\tau_1$, $\tau_2$ and $\tau_3$
drops rapidly to zero. This is a clear signature of the localized states, whose localization length is much smaller than
$t_0$. In contrast, $\log_2(P_4)$ and $\log_2(P_5)$ increase almost linearly as a function of $\log_2(t_l)$. The
corresponding exponent $\tau_4$ and $\tau_5$ are always larger than $0.5$ for $t_l<t_0$. This indicates that their
localization length is larger than $t_0$, which agrees with the intuitive picture obtained from Fig.~\ref{fig5}(a) and
(b). It suggests that the wave functions are essentially delocalized, which are confined by the two pulses. This is in
analogy with the "particle in a box" problem in one-dimensional system.

\begin{figure}
  \centering
  \includegraphics{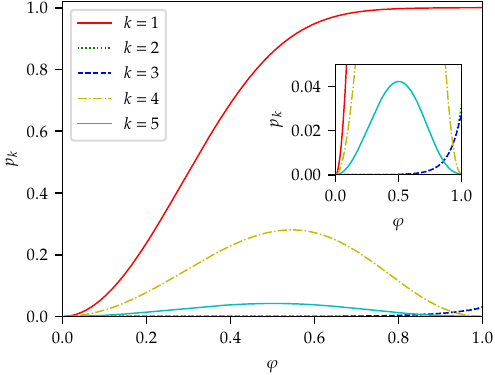}  
  \caption{ The excitation probabilities $p_k$ as a function of $\varphi$ for $\varphi = 0.9$ and $t_0=32$. The inset
    shows the zoom-in for $p_k < 0.05$. The probabilities $p_k$ of other \textit{eh} pairs are too small to be seen from
    the figure.}
  \label{fig-s3-30}  
\end{figure}

Having identified the presence of delocalized states, now we answer the question how does these states emerge as the
flux $\varphi$ decreases from $1.0$. To show this, we first plot the excitation probabilities as a function of $\varphi$
in Fig.~\ref{fig-s3-30}, corresponding to $t_0=32$ and $\varphi \in [0, 1]$. While all the probabilities eventually
drops to zero as $\varphi$ approaches $0.0$. They exhibit three different dependences on $\varphi$: 1) $p_2$ and $p_3$
are almost degenerated, which drop monotonically as $\varphi$ decreases and vanish for $\varphi<0.5$. This can be seen
more clearly from the zoom-in shown in the inset. 2) $p_4$ and $p_5$ exhibit a non-monotonic dependence on $\varphi$,
which take their maximum value when $\varphi$ is slightly larger than $0.5$ and vanish for $\varphi=1.0$. 3) $p_1$ drops
monotonically from $1.0$ to $0.0$.

\begin{figure}
  \centering
  \includegraphics{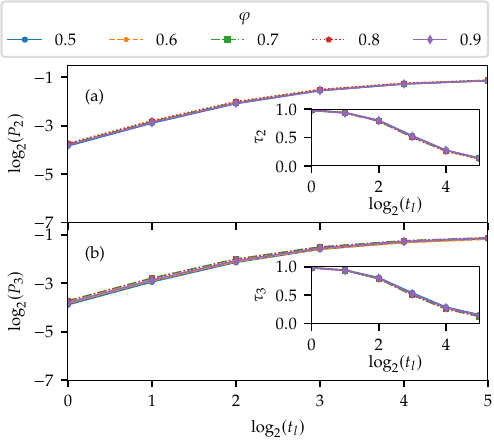}
  \caption{ The logarithm of the IPR $\log_2(P_k)$ as a function of logarithm of the box size $\log_2(t_l)$ for $t_0=32$
    and $\varphi=0.9$, where (a) corresponding to $k=2$ and (b) corresponding to $k=3$. The exponent $\tau_2$ and
    $\tau_3$ estimated from the slope of the logarithm plot via Eq.~(\ref{s3:eq30}) are demonstrated in the insets.}
  \label{fig-s3-40}
\end{figure}

The degeneracy of $p_2$ and $p_3$ indicates that they correspond to the neutral \textit{eh} pairs whose wave functions
are bonding and anti-bonding states built from localized states. The localization nature can be seen from the
$t_l$-dependence of their IPRs, which are shown in Fig.~\ref{fig-s3-40}(a) and (b). In the main panel of the figures,
curves with different markers and line types correspond to IPRs with different values of $\varphi$. The corresponding
exponents $\tau_2$ and $\tau_3$ estimated from the slope via Eq.~(\ref{s3:eq30}) are shown in the insets. Note that we
only show probabilities for $\varphi>0.5$ when $p_2$ and $p_3$ is not too small. One can see that all IPRs are
insensitive to $\varphi$. They all approach constants as $t_l$ reaches $t_0$ and the corresponding exponents $\tau_2$
and $\tau_3$ drop to zero. This indicates that the corresponding wave functions remain localized as $\varphi$ varies.

\begin{figure}
  \centering
  \includegraphics{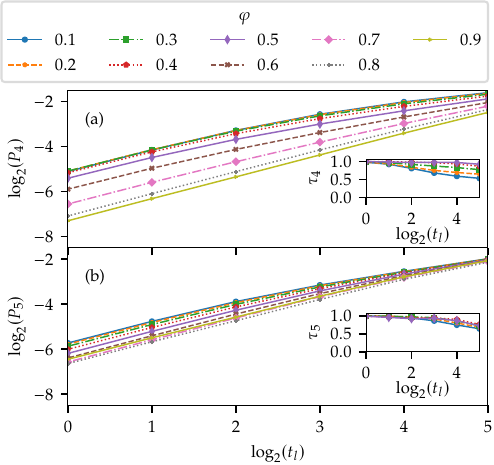}
  \caption{ The same as Fig.~\ref{fig-s3-40} but for the delocalized neutral \textit{eh} pairs with $k=4$ (a) and $k=5$
    (b).}
  \label{fig-s3-50}
\end{figure}

In contrast, probabilities $p_4$ and $p_5$ correspond to the delocalized neural \textit{eh} pairs. The $t_l$-dependence
of their IPRs for typical values of $\varphi$ are shown in Fig.~\ref{fig-s3-50}(a) and (b), while the exponent $\tau_4$
and $\tau_5$ estimated from the slopes are shown in the inset. Comparing to Fig.~\ref{fig-s3-40}, the IPRs for the two
delocalized states are more sensitive to $\varphi$. But the exponents $\tau_4$ and $\tau_5$ are always larger than
$0.5$. This indicates that the corresponding wave functions remain delocalized as $\varphi$ varies. Hence the excitation
of the neutral \textit{eh} pairs are dominated by the delocalized ones when $\varphi$ is sufficiently small. This is
induced by the enhancement/suppression of the excitation probabilities for the delocalized/localized \textit{eh}
pairs. Their wave functions remain localized and/or delocalized as $\varphi$ varies.

\begin{figure}
  \centering
  \includegraphics{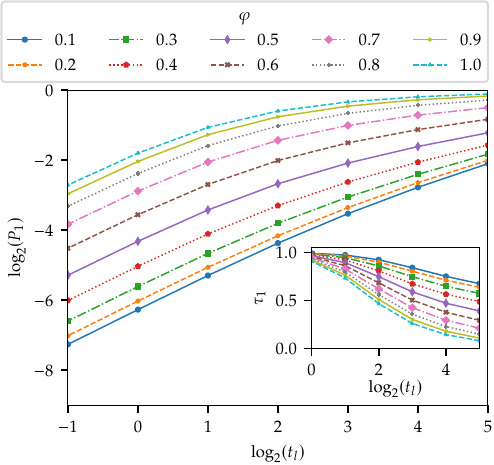}
  \caption{The same as Fig.~\ref{fig-s3-40} but for the individual electron with $k=1$.}
  \label{fig-s3-60}
\end{figure}

However, this is not the case for individual electrons and holes, corresponding to the excitation probability $p_1$. To
show this, we plot the corresponding IPR $P_1$ as a function of $t_l$ in Fig.~\ref{fig-s3-60} for different typical
value of $\varphi$. The corresponding exponent $\tau_1$ estimated from the slope are shown in the inset. One can see
that the $t_l$-dependence of the IPR shows a qualitatively different behavior from the neutral \textit{eh} pairs as
$\varphi$ drops from $1.0$ to $0.0$. For $\varphi=1.0$, the IPR tends to be $t_l$-independent as $t_l$ approaches
$t_0$. The corresponding exponent $\tau_1$ for $t_l=t_0$ is $0.08$, which is very close to zero. This indicates that it
corresponds to a localized state, whose localization length is much smaller than $t_0$. In contrast, the IPR as a
function of $t_l$ increases linearly on log-log scale for $\varphi=0.1$. The corresponding exponent $\tau_1$ for
$t_l=t_0$ is $0.68$, which is apparently larger than zero. This indicates that corresponding localization length is much
larger than $t_0$. Hence one expect the wave function can undergo a LD transition as $\varphi$ drops from $1.0$ to
$0.0$.

Due to the effect of the finite system size, the presence of the LD transition cannot be confirmed simply from
Fig.~\ref{fig-s3-60}. A detailed finite-size scaling analysis is required, which shall be performed in the following
section.

\section{Single-parameter Scaling of the LD transition}
\label{sec4}

Following the single-parameter scaling theory \cite{soukoulis_1984_fract, schreiber_1985_fract, castellani_1986_multif,
  bauer_1990_correl, schreiber_1991_multif, yakubo_1998_finit, iyer_2013_many, rodriguez_2011_multif,
  tikhonov_2016_ander, puschmann_2021_quart}, we assume the IPR $P_1$ for different $t_l$, $t_0$ and $\varphi$ can be
described by an universal scaling function as
\begin{equation}
  P_1(t_l, t_0, \varphi) = \xi^{\eta} f[t_l/\xi(\varphi), t_0/\xi(\varphi)],
  \label{s4:eq10}
\end{equation}
where $\eta$ is a constant exponent, $\xi(\varphi)$ is the correlation length of infinite system and $f(x, y)$ is the
scaling function. If the LD transition occurs, the correlation length $\xi(\varphi)$ shall diverge at a critical value
$\varphi_c$.

To obtain the correlation length $\xi$, we adopt the data collapse method. We first calculate the IPR $P_1$ for various
$t_l$ and $\varphi$ for a given $t_0$. Then we rescale the IPR by choosing the parameters $\eta$ and $\xi$ so that all
the data points can be collapsed into a single curve, which corresponds to the scaling function for a fixed $t_0$. The
parameters $\eta$ and $\xi$ can be further constrained by noting that $P_1$ tends to be independent on $t_l$ in the
localized limit. Without loss of generality, we assume the scaling function approaches a constant in this limit, which
implies $\eta=0$. This provides the data collapse method with only one free parameter $\xi$ for a given $t_0$, which can
be performed numerically without priori assumption of the unknown scaling function \cite{mackinnon_1981_one}. Finally,
we compare $\xi$ as a function of $\varphi$ for different $t_0$ to check the impact of $t_0$. The correlation length
$\xi(\varphi)$ is expected to be independent on $t_0$ when $t_0$ is large enough.

\begin{figure}
  \centering
  \includegraphics{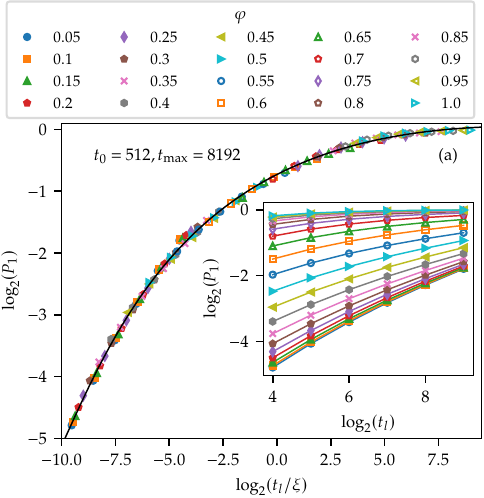}
  \includegraphics{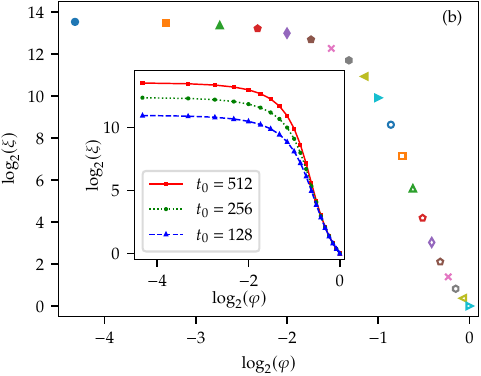}  
  \caption{ (a) The IPR $\log_2(P_1)$ as a function of $\log_2(t_l/\xi)$, corresponding to a pair of Gaussian pulse with
    $t_0=512$ and $t_{\rm max}=8192$. The IPR before the rescaling are shown in the inset. (b) The correlation length
    $\log_2(\xi)$ as a function of $\log_2(\varphi)$, corresponding to a pair of Gaussian pulse with $t_0=512$ and
    $t_{\rm max}=8192$. The inset shows the correlation length for $t_0=256$ and $t_0=512$, while all other parameters
    remaining the same. }
  \label{fig11}
\end{figure}

The collapse of the IPR as a function of $t_l$ is demonstrated on log-log scale in the main panel of
Fig.~\ref{fig11}(a), corresponding to a pair of Gaussian pulses with $t_0=512$. In the numerical calculation, we
restrict ourself in the time domain $t \in [-t_{\rm max}/2, +t_{\rm max}/2]$ with $t_{\rm max}=8192$ so that it provide
a compact support of all the wave functions. The inset shows the IPR before the rescaling. Curves with different makers
correspond to different value of $\varphi$. By properly chosen $\xi$, we find that the rescaled IPR can be
well-collapsed into a single curve, as illustrated by the black solid curve in the main panel.

The corresponding correlation length $\xi$ is plotted on log-log scale in the main panel of Fig.~\ref{fig11}(b). In the
figure, data points with different markers represent $\xi$ for different value of $\varphi$. In the data collapse
method, the value of $\xi$ is only determined up to a constant multiplicative factor. To fix this factor, we assume the
correlation length is finite in the localized limit, which is comparable to the width of the voltage pulse. We simply
choose $\xi=1.0$ for $\varphi=1$ in the figure, which is expected to be a good estimation. From the figure, one can see
that $\xi$ diverges rapidly as $\varphi$ decreases from $1.0$ to $0.0$, this is a key signature of the phase transition,
corresponding to the critical value $\varphi_c=0.0$. The divergence is rounded when $\xi$ is large, which can be
attributed to the finite-size effect. The finite-size effect can be better seen from the inset of Fig.~\ref{fig11}(b),
where we compare the correlation length as a function of $\varphi$ for $t_0=128$, $256$ and $512$, while all other
parameters remaining the same. One can see that the correlation length is insensitive to $t_0$ when $\xi$ is small, but
becomes sensitive to $t_0$ when $\xi$ is sufficiently large. The above results show that the wave function of individual
electron (hole) does exhibit a LD transition as $\varphi$ decreases from $1.0$ to $0.0$. This is distinctly from the
wave function of the neutral \textit{eh} pairs.

\begin{table}
  \caption{Voltage pulses}
  \begin{ruledtabular}
    \begin{tabular}{ll}
      Name & Expression \\
      Gaussian & $V_p(t) = \frac{\varphi}{\sqrt{\pi\ln(2)}} \exp[ -\ln(2)t^2 ]$\\
      Hyperbolic secant & $V_p(t) = \frac{\varphi}{\pi} \ln(2+\sqrt{3}) \sech[\ln(2+\sqrt{3})t]$\\
      Lorentzian-squared & $V_p(t) = \frac{\varphi}{\pi^2} \frac{1}{\left( t^2+1 \right)^2}$\\      
      Lorentzian & $V_p(t) = \frac{\varphi}{2\pi^2} \frac{1}{t^2+1}$\\      
    \end{tabular}
  \end{ruledtabular}
  \label{s4:tab10}
\end{table}

Up to now, we have only focused on the Gaussian pulses. One may thus wonder if the LD transition is sensitive to the
detailed profile of the pulse. To see this, we compare the LD transition for four different voltage pulses as shown in
Table~\ref{s4:tab10}. We choose these pulses because they exhibit different long-time tails, as the phase transition is
usually dominated by low energy behaviors.

\begin{figure}
  \centering
  \includegraphics{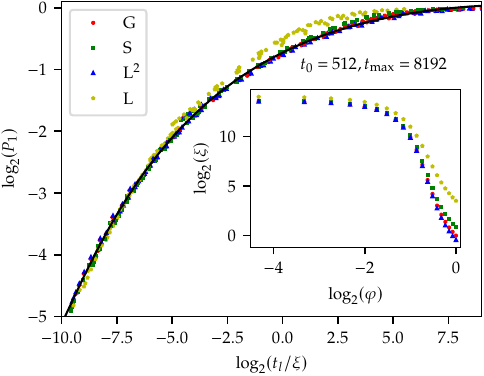}
  \caption{ The IPR $\log_2(P_1)$ as a function of $\log_2(t_l/\xi)$ for $t_0=512$ and $t_{\rm max}=8192$. The red
    circles, green squares, blue triangles and yellow pentagons correspond to the Gaussian (G), Hyperbolic secant (S),
    Lorentzian-squared (${\rm L}^2$) and Lorentzian (L) pulses. The corresponding correlation length $\log_2(\xi)$ as a
    function of $\log_2(\varphi)$ is shown in the inset.}
  \label{fig12}
\end{figure}

The collapse of the IPR as a function of $t_l$ for the for different voltage pulses are demonstrated on log-log scale in
the main panel of Fig.~\ref{fig12}, corresponding to $t_0=512$ and $t_{\rm max}=8192$. By properly choosing $\xi$, we
find that the IPR for the Gaussian (G), Hyperbolic secant (S) and Lorentzian-squared (${\rm L}^2$) pulses can be
well-collapsed into the same scaling function, as illustrated by the black solid curve in the figure. In the meantime,
the corresponding correlation lengths also diverge in a similar manner, as illustrated in the inset. In contrast, the
IPR collapse into a different scaling function for the Lorentzian (L) pulse, as shown by the yellow pentagons in the
main panel. The corresponding correlation length also exhibit a quantitatively different dependence on $\varphi$, as
illustrated in the inset. This suggests that the corresponding LD transition belongs to a different universality class.

\section{Summary and Outlook}
\label{sec5}

In summary, we have studied the wave function of the individual electrons (holes) and neutral \textit{eh} pairs injected
by voltage pulses with non-integer flux quantum. We have shown that the wave function can be delocalized in the time
domain, which can be measured by using the IPR. We found that the wave function exhibit different dependence on as
$\varphi$ drops from $1.0$ to $0.0$. The wave functions of the neutral \textit{eh} pairs remain delocalized or localized
as $\varphi$ varies. In contrast, the wave functions of individual electrons (holes) undergo a LD transition, which can
be described by using a single-parameter scaling theory. We found that the LD transition is universal for three typical
profiles of voltage pulses (Gaussian, Hyperbolic secant and Lorentzian-squared), which can be described by the same
scaling function and correlation length. In contrast, the LD transition for the Lorentzian pulse exhibit a different
scaling function and correlation length, indicating that it belongs to a different universality class.


\bibliographystyle{apsrev4-2}
%

\end{document}